\documentclass[prc,preprint,showpacs,preprintnumbers,amsmath,amssymb,superscriptaddress]{revtex4-1}
\usepackage{graphicx}
\usepackage{dcolumn}
\usepackage{bm}
\usepackage{floatrow}
\usepackage[caption=false]{subfig}
\usepackage{booktabs}
\usepackage{multirow}
\usepackage{siunitx}
\usepackage{subfig}
\usepackage{longtable}
\usepackage{footnote}

\begin{document}
\preprint{}

\title{Multiparticle-hole excitations in nuclei near N = Z = 20 : $^{41}$K}

\author{E. Rubino}
\altaffiliation{Present address: NSCL, MSU, East Lansing, Michigan 48824, USA.}
\affiliation{Department of Physics, Florida State University, Tallahassee, Florida 32306, USA.}
\author{S. L. Tabor} 
\affiliation{Department of Physics, Florida State University, Tallahassee, Florida 32306, USA.}
\author{Vandana Tripathi}
\affiliation{Department of Physics, Florida State University, Tallahassee, Florida 32306, USA.}
\author{R. S. Lubna}
\altaffiliation{Present address: NSCL, MSU, East Lansing, Michigan 48824, USA.}
\affiliation{Department of Physics, Florida State University, Tallahassee, Florida 32306, USA.}
\author{B. Abromeit}
\altaffiliation{PNNL, Richland, Washington 99354, USA.}
\affiliation{Department of Physics, Florida State University, Tallahassee, Florida 32306, USA.}
\author{J. M. Allmond}
\affiliation{Physics Division, Oak Ridge National Laboratory, Oak Ridge, Tennessee 37831, USA.}
\author{L.T. Baby}
\affiliation{Department of Physics, Florida State University, Tallahassee, Florida 32306, USA.}
\author{D.D. Caussyn}
\affiliation{Department of Physics, Florida State University, Tallahassee, Florida 32306, USA.}
\author{K. Kravvaris}
\affiliation{Department of Physics, Florida State University, Tallahassee, Florida 32306, USA.}
\altaffiliation{Lawrence Livermore National Laboratory, Livermore, CA 94550, USA.}
\author{A. Volya}
\affiliation{Department of Physics, Florida State University, Tallahassee, Florida 32306, USA.}

\date{\today} 

\begin{abstract}

This experimental study of high-spin structure near N = Z = 20 nuclei was focused on $^{41}$K, but will also mention three newly observed $\gamma$ transitions in $^{41}$Ca observed in the same reaction. High-spin states were populated using the $^{26}$Mg($^{18}$O, $p2n\gamma$)$^{41}$K and $^{26}$Mg($^{18}$O, $3n\gamma$)$^{41}$Ca reactions. The experiment was carried out at an incident beam energy of 50 MeV at the Florida State University (FSU) John D. Fox Superconducting Linear Accelerator Laboratory and used the FSU high-purity germanium detector array. The $^{41}$K level scheme was extended to 12325 keV, possibly with J$^{\pi}$ = 25/2$^-$ or 27/2$^+$, by means of 25 new transitions and that of $^{41}$Ca to 9916 keV. Linear polarization and a measure of angular distribution results are also reported and used to provide information on the spins and parities of several states in the $^{41}$K decay scheme. The results have been compared to the $spsdpf$ cross-shell FSU shell model interaction calculations. The theoretical results from configurations involving no or one additional nucleon promoted from the $sd$ to the $fp$ shell agree relatively well with the energies of known states, while those that involve multi-particle excitations paint an interesting and complex picture of interplay between single-particle excitations, collective pairing, and deformation. This presents an interesting challenge for future theory.

\end{abstract}

\maketitle 
 
%
%
\section{Introduction}
The concept of shells, and the gaps between them, has proved to be a very insightful tool in nuclear structure since the original proposal by Maria Goeppert Mayer \cite{meyer1948}, inspired by the N = Z = 20 shell gap \cite{wigner1937,barkas1939}. Though the shell model has been quite successful generally, understanding the shell structure in the A = 40 region has been challenging. Higher-spin states further explore cross-shell excitations which play a critical role around shell boundaries.

These considerations led us to perform a fusion-evaporation reaction to further explore the high-spin structure of $^{41}$K (Z = 19, N = 22) and $^{41}$Ca (Z = 20, N = 21) which straddle the $sd$-$fp$ shell boundary. Fusion reactions between comparable mass projectile and target produce compound nuclei with the highest possible angular momenta. The evaporation of light particles does not remove much of the angular momentum, therefore it is dissipated through $\gamma$ decay. Due to the preference of electromagnetic decay to change spin by only 1 or 2 with the highest decay energy possible, the resulting $\gamma$ decays tend to cascade through the lowest states of a given spin which are called yrast states. The highest spins achievable at a given excitation energy tend to be unique and can be more reliably compared to theoretical structure calculations. Additionally, our group has recently developed a new $spsdpf$ cross-shell interaction. It has been quite successful in predicting higher-spin cross-shell excitations in nuclei a little lighter than $A \approx 40$ and is compared to our experimental results further on \cite{PhysRevC.100.014310,lubna_structure_2019,lubna_evolution_2020}. 

Since this project is focused on the role of excitations across the N = Z = 20 boundary, it is important to define our nomenclature. The 0 particle - 0 hole (0ph) configurations do not involve any additional nucleons crossing the shell gap relative to the predominant ground state (g.s.) configuration. The 1ph excitations involve the promotion of 1 nucleon (either 1 proton or 1 neutron or any combination adding up to 1) from the $1s0d$ shell to the $0f1p$ one. Such states will have an opposite parity to that of the g.s. In fact all even (odd) ph excitations will have the same (opposite) parity as the g.s. This also provides an experimental signature of the nucleus' structure if the parity of the state can be determined.

First, in this section we will summarize the previously reported $\gamma$ spectroscopy results on higher-spin states in $^{41}$K and $^{41}$Ca. An extensive study of $\gamma$-ray polarizations following a variety of fusion-evaporation reactions, including $^{18}$O + $^{26}$Mg, using a 2-crystal Ge(Li) Compton polarimeter was able to identify states up to about 3.9 MeV excitation energy with a suggested spin-parity of 15/2$^+$ and about 5 MeV with a possible spin of 19/2 in the $^{41}$K decay scheme \cite{BNL}. The same reaction at 34 MeV was used by another group employing a Ge(Li)-NaI(Tl) Compton suppression spectrometer and a three-crystal Ge(Li) Compton polarimeter. This work was able to confirm the spin and parity assignments of many states \cite{Utrecht}. A third experiment with the same reaction focused on the measurement of level lifetimes using the Doppler-shift attenuation method (DSAM) \cite{Koln}.

Several of these experiments also reported on $^{41}$Ca. States up to 6826 keV, J$^{\pi}$ = (19/2)$^-$ were identified in Ref. \cite{BNL}. The $^{41}$Ca nucleus was most recently studied using a $^{27}$Al($^{16}$O,$pn$)$^{41}$Ca fusion-evaporation reaction with a 34 MeV beam. The resulting $\gamma$ de-exciting transitions were detected using the Indian National Gamma Array (INGA) composed of 18 Compton-suppressed high-purity germanium (HPGe) clover spectrometers. This work extended the $^{41}$Ca decay scheme to $\sim$9 MeV. Spin and parity assignments were made using coincidence intensity anisotropies and linear polarization measurements. Additionally, lifetimes were measured using the DSAM \cite{NNDC} . 


With these thoughts in mind, we studied the $^{18}$O + $^{26}$Mg fusion-evaporation reaction at a higher beam energy of 50 MeV using a larger $\gamma$ detection array than had been used in any of the previous $^{41}$K experiments. For the $^{41}$Ca studies, the efficiency of our array did not exceed that of INGA, but we were surprisingly able to detect three newly observed $\gamma$ rays. Additionally, the sensitivity to higher-spin states resulted from the use of a thin target. We were also able to verify many of the previously observed transitions and states. 
%
%
\section{Experiment}
\label{exp}
The $^{26}$Mg($^{18}$O,p2n)$^{41}$K and $^{26}$Mg($^{18}$O,3n)$^{41}$Ca reaction channels were studied at the FSU John Fox Lab to explore high-spin, $sd$-$fp$ cross-shell states. The $^{18}$O beam was accelerated to 50 MeV using the 9 MV Super-FN tandem Van de Graaff accelerator. The self-supporting $^{26}$Mg target was 770-$\mu$g/cm$^2$ thick and 99.6\% isotopically enriched. A 25 micron tantalum stopping foil was used to stop the beam to avoid irradiating the 0$^o$ silicon detectors but still allow the light charged particles to pass through with fairly small energy loss.

Two experiments were carried out using similar setups as well as the same beam and target. The bulk of this analysis was performed on the one that used six HPGe clover detectors and two single-crystal HPGe detectors. Of these, three HPGe clover detectors were Compton suppressed with bismuth germanate (BGO) shields as well as one single-crystal HPGe detector. Two of the HPGe clover detectors were placed at 135$^o$, the other four HPGe clover detectors were placed at 90$^o$, one single-crystal HPGe detector was also placed at 90$^o$, the last single-crystal HPGe detector was placed at 45$^o$. An E-$\Delta$E silicon particle telescope was placed at 0$^o$ to detect the light charged particles from the evaporation. The E detector was 1.0 mm thick and the $\Delta$E detector, 0.1 mm thick.

The HPGe energy and efficiency calibrations were carried out using a combination of the radioactive sources: a calibrated $^{152}$Eu source from NIST\cite{NIST}, $^{133}$Ba, and $^{60}$Co. An effective $\beta$ (= $v$/$c$) of 0.028 was used to correct the Doppler shifting of $\gamma$ rays in the spectra.

The data acquisition system consisted of a Digital Gamma Finder Pixie16 system \cite{Pixie} with a sampling rate of 100 MHz. The minimum trigger condition was either two HPGe detectors or one HPGe detector and the $\Delta$E detector. The data were analyzed using GNUSCOPE \cite{gnuscope}. The data were sorted into several matrices: a Doppler corrected-not Doppler corrected square matrix; a Doppler corrected triangle matrix; a non-Doppler corrected triangle matrix all for $\gamma$-$\gamma$ coincidences; and a Doppler corrected, proton-gated triangle matrix for proton-$\gamma$-$\gamma$ coincidences. The gain-corrected energies of multiple coincident hits of crystals in a Clover spectrometer were added together (add-back) to improve the statistics of higher energy transitions and reduce the Compton scattered background at lower energies. Add-back was not used for the polarization analysis.

The polarization of radiation from an aligned nucleus provides information on the parity change of the decay. We measured polarizations relative to the beam axis using the sensitivity of Compton scattering between pairs of crystals in the Clover detectors placed approximately perpendicular to the $^{18}$O beam \cite{starosta_experimental_1999}. Only those events in which there were signals in coincidence from exactly two crystals were examined (parallel or perpendicular to the beam). These pairs of energy signals were added after gain matching and Doppler correction and were observed to produce spectra comparable to the single crystal ones, albeit at a substantially reduced intensity. The spectra of all the pairs parallel (perpendicular) to the beam were added together for the clover spectrometers placed near 90$^0$. In this arrangement perpendicular (parallel) Compton scattering was somewhat stronger for predominantly electric (magnetic) radiation \cite{KleinNishina}. The differences are small and were multiplied by 10 in the figures for ease of comparison. To better select $^{41}$K or $^{41}$Ca lines from the dominant $^{38}$Ar ones, coincidences with any one of their strong low-lying transitions at any angle were required.

A measure of the angular distribution \cite{RevModPhys.31.711} of many $^{41}$K transitions are listed in the table discussed further on. This measure is the ratio of peak areas from the detectors at $135^o$ and $45^o$ to that of those at $90^o$ normalized to the ratio measured in the stronger lines from the $^{152}$Eu source which must be isotropic. Two methods were used to reduce contaminant lines. One was to measure these ratios in the smaller proton-gated data set for $^{41}$K lines. The other was to gate the lines in $\gamma$-$\gamma$ matrices of the selected angle(s) on one axis with all detectors on the other axis. To increase statistics and reduce correlation effects, all the stronger lines in the same decay sequence were used as gates. Where possible, ratios determined by the two methods were averaged, with the difference providing part of the estimate of the uncertainty. These ratios, R$_{\rm AD}$, were used in a standard angular distribution code to test how they fit various spin hypotheses and for what mixing ratios $\delta$.

%
%
\section{Results}
\subsection{$^{41}$K}

\begin{figure}[h!]
\includegraphics[width=\linewidth]{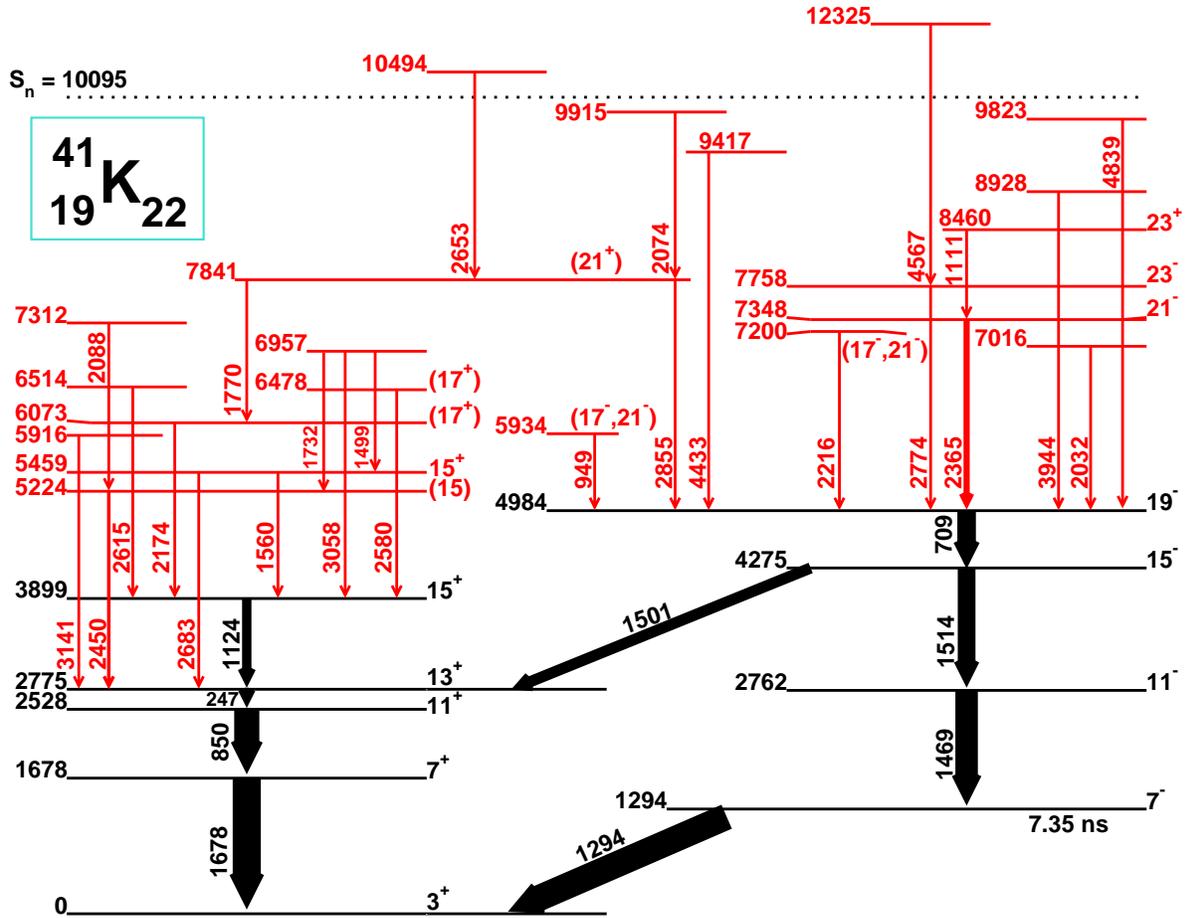}	
\caption{States (with spins given as 2J) and $\gamma$ decays observed in the present experiment in $^{41}$K (with newly observed ones labelled in red). The neutron separation energy, S$_n$, is represented by a horizontal dotted line. The newly observed states are placed above the previously known decay sequences to which they decay. This does not imply that they have the same parities as the states they decay to, in fact the tentative spin-parity assignments are discussed in the text. The widths of the decay arrows give an indication of their intensities. Note that the lifetime of the 1294 keV state comes from Ref. \cite{NNDC}.}

\label{fig:41K_lvl}
\end{figure}

\begin{figure}[h!]
\includegraphics[width=\linewidth]{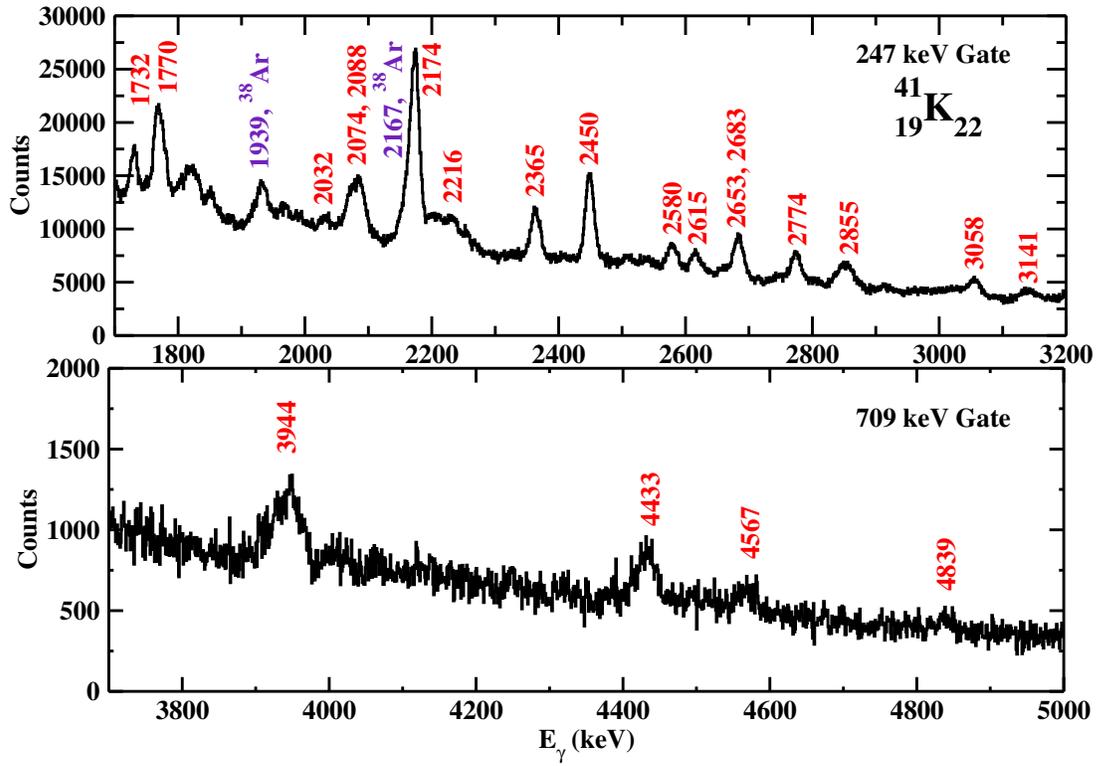}	
\caption{Portions of the $\gamma$ spectra in coincidence with the 247 keV (top) and 709 keV (bottom) transitions showing many of the newly observed $\gamma$ lines (in red). The $^{38}$Ar contaminant transitions are labeled in purple.}
\label{fig:709}
\end{figure}

\begin{figure}[h!]
\includegraphics[width=\linewidth]{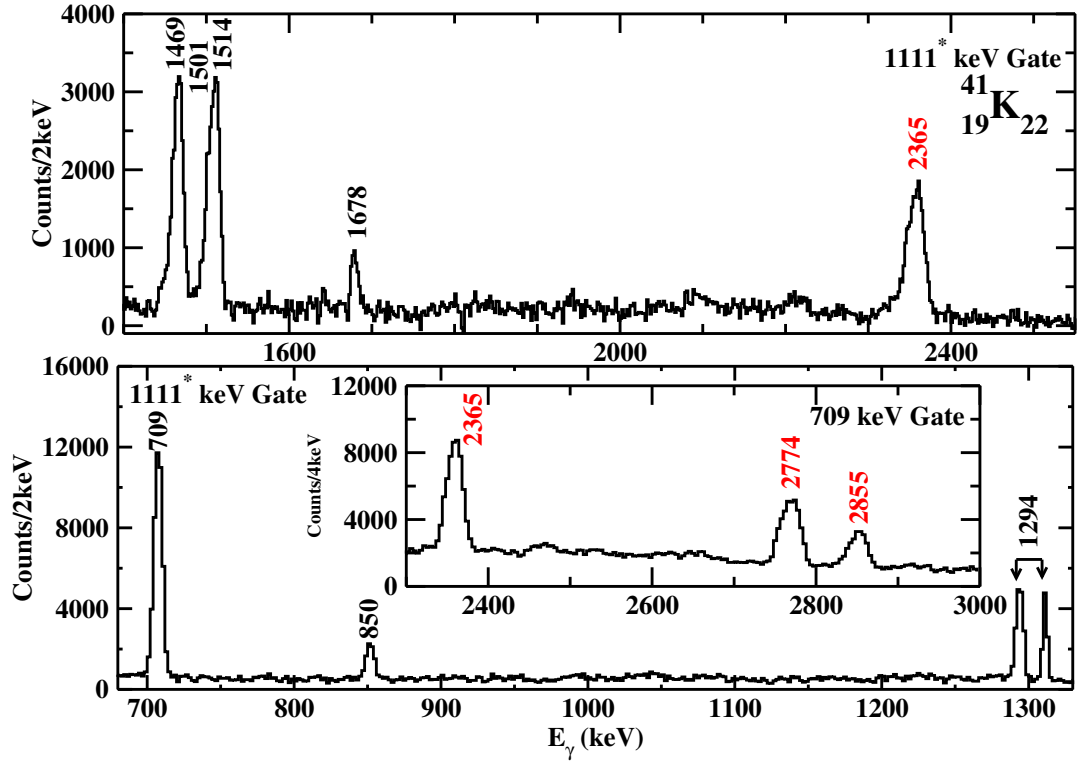}	
\caption{A portion of the $\gamma$ spectrum in coincidence with the newly observed 1111 keV $\gamma$ transition showing mostly the previously known lines in $^{41}$K. An extension of the 709 keV coincidence spectrum showing newly observed higher-energy lines (in red) is shown in the inset.}
\label{fig:1111}
\end{figure}

The decay scheme consisting of transitions and levels observed with this experiment (newly observed ones are labeled in red) is shown in Fig. \ref{fig:41K_lvl}. Approximate relative intensities of the transitions are represented by varying arrow thickness from the efficiency-corrected areas of these peaks in the total projection spectrum. This analysis resulted in the addition of 21 new states, including two above the neutron separation energy, and 25 new $\gamma$ transitions in the $^{41}$K decay scheme extending it to 12325 keV. The coincidence spectra in Figs. \ref{fig:709} and \ref{fig:1111} are those with the 247 keV, 709 keV (the main positive and negative parity sequences, respectively), and 1111 keV $\gamma$ transitions and show several of the newly observed transitions. The excitation energies, transition energies, dominant radiation mode, angular distribution ratios R$_{\rm AD}$, and initial and final J$^{\pi}$ values are given in Table \ref{Tab:41K}.

\clearpage
\newpage
\begin{center}
\begin{longtable}[h]{ccccccc} 
\caption{\label{Tab:41K}{Energies of newly observed states and transitions (denoted by $^*$) as well as the levels fed by those transitions in $^{41}$K. Previously known J$^{\pi}$ values and current suggestions are shown along with the decay modes based on polarization analyses and R$_{AD}$ values.}}\\
 \hline
\multicolumn{1} {c}{E$_{x,i}$ (keV)} & \multicolumn{1} {c}{E$_{x,f}$ (keV)} & \multicolumn{1} {c}{E$_\gamma$ (keV)} & \multicolumn{1} {c}{Decay Mode} & \multicolumn{1} {c}{2J$^\pi_i$} & \multicolumn{1} {c}{2J$^\pi_f$} & \multicolumn{1} {c}{R$_{AD}$}\\ 
\hline\hline
\endfirsthead

\multicolumn{4}{c}
{{\bfseries \tablename\ \thetable{} -- continued from previous page}} \\
\hline
\multicolumn{1} {c}{E$_{x,i}$ (keV)} & \multicolumn{1} {c}{E$_{x,f}$ (keV)} & \multicolumn{1} {c}{E$_\gamma$ (keV)} & \multicolumn{1} {c}{Decay Mode} & \multicolumn{1} {c}{2J$^\pi_i$} & \multicolumn{1} {c}{2J$^\pi_f$} & \multicolumn{1} {c}{R$_{AD}$}\\
\hline\hline
\endhead

\hline \multicolumn{4}{r}{{ Continued on next page}} \\ \hline
\endfoot

\hline \hline
\endlastfoot
1293.6 (3) & 0 & 1293.6 (3) & M & 7$^-$ & 3$^+$ & \\

1677.6 (4) & 0 & 1677.6 (4) & E & 7$^+$ & 3$^+$ & 1.31 (9)\\

2528.0 (5) & 1677.6 & 850.4 (5) & E & 11$^+$ & 7$^+$ & 1.25 (4)\\

2762.4 (8) & 1293.6 & 1468.8 (7) & E & 11$^-$ & 7$^-$ & 1.36 (7)\\

2774.8 (7) & 2528.0 & 246.8 (3) & M & 13$^+$ & 11$^+$ & 0.87 (6)\\ 

3898.6 (16) & 2774.8 & 1123.8 (10) & M & 15$^+$ & 13$^+$ & 1.02 (3)\\

4275.4 (15) & 2762.4 & 1514.2 (14) & E & 15$^-$ & 11$^-$ & 1.39 (3)\\
 & 2774.8 & 1500.6 (14) & E & 15$^-$ & 13$^+$ & \\

4983.9 (15) & 4275.4 & 708.6 (3) & E & 19$^-$ & 15$^-$ & 1.31 (3)\\ 

5224.2$^*$ (7) & 2774.8 & 2449.5$^*$ (3) & & (15) & 13$^+$ & 0.93 (16)\\

5458.7$^*$ (16) & 2774.8 & 2683.2$^*$ (7) & (M) & 15$^+$ & 13$^+$ & 1.16 (14)\\ 
 & 3898.6 & 1560.2$^*$ (3) & & 15$^+$ & 15$^+$ & \\ 

5915.9$^*$ (7) & 2774.8 & 3141.1$^*$ (7) & & & 13$^+$ & \\

5933.8$^*$ (15) & 4983.9 & 949.4$^*$ (3) & (M) & (17$^-$, 21$^-$) & 19$^-$ & 1.14 (10)\\

6072.5$^*$ (16) & 3898.6 & 2174.0$^*$ (3) & & (17$^+$) & 15$^+$ & 0.85 (10)\\

6478.4$^*$ (16) & 3898.6 & 2579.9$^*$ (3) & & (17$^+$) & 15$^+$ & 0.58 (25)\\ 

6513.6$^*$ (16) & 3898.6 & 2615.1$^*$ (3) & & & 15$^+$ & \\

6956.6$^*$ (16) & 3898.6 & 3057.6$^*$ (7) & & & 15$^+$ & \\ 
 & 5224.2 & 1731.5$^*$ (4) & & & (15) & \\ 
 & 5458.7 & 1499.4$^*$ (15) & & & 15$^+$ & \\

7016.0$^*$ (15) & 4983.9 & 2032.1$^*$ (3) & & & 19$^-$ & \\

7199.5$^*$ (15) & 4983.9 & 2215.6$^*$ (3) & M & (17$^-$, 21$^-$) & 19$^-$ & 0.59 (20)\\

7312.2$^*$ (15) & 5224.2 & 2088.0$^*$ (3) & & & (15) & \\

7348.4$^*$ (15) & 4983.9 & 2364.5$^*$ (3)& M & 21$^-$ & 19$^-$ & 1.59 (20)\\ 

7757.9$^*$ (15) & 4983.9 & 2774.0$^*$ (3) & (E) & 23$^-$ & 19$^-$ & 1.54 (20)\\

7840.9$^*$ (16) & 4983.9 & 2855.4$^*$ (6) & (M) & (21$^+$) & 19$^-$ & \\
 & 6072.5 & 1770.0$^*$ (3) & (E) & (21$^+$) & (17$^+$) & \\

8459.5$^*$ (15) & 7348.4 & 1111.1$^*$ (3)& E & 23$^+$ & 21$^-$ & 0.73 (20) \\

8927.6$^*$ (25) & 4983.9 & 3943.7$^*$ (20) & & & 19$^-$ & \\

9416.6$^*$ (16) & 4983.9 & 4432.7$^*$ (15) & & & 19$^-$ & \\

9822.6$^*$ (21) & 4983.9 & 4838.7$^*$ (15) & & & 19$^-$ & \\

9914.5$^*$ (16) & 7840.9 & 2073.6$^*$ (3) & & & (21$^+$) & \\

10493.6$^*$ (19) & 7840.9 & 2652.7$^*$ (10) & & & (21$^+$) & \\

12325.2$^*$ (21) & 7757.9 & 4567.3$^*$ (15) & & & 23$^-$ & \\
\end{longtable}
\end{center}

\begin{figure}[h!]
\includegraphics[width=\linewidth]{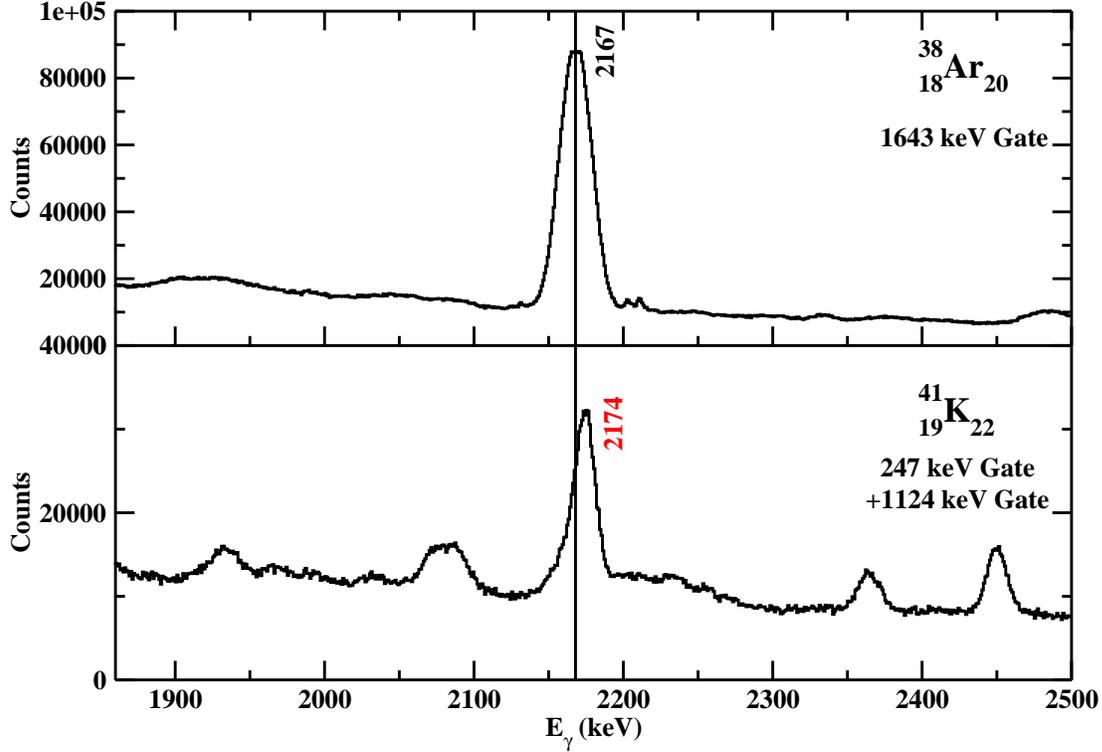}	
\caption{Transitions in coincidence with the $^{38}$Ar 1643 keV transition (denoted in purple), in particular the 2167 keV transition, are shown in the top graph and those in coincidence with the $^{41}$K 247 and 1124 keV transitions, in particular the newly observed 2174 keV transition, are shown in the bottom graph, illustrating the difference in peak positions.}
\label{fig:2174}
\end{figure}

The largest reaction product was $^{38}$Ar, we have been careful to discriminate those transitions from the proposed new ones in $^{41}$K. An important case is the new 2174 keV line which lies uncomfortably close to the 2167 keV $2^+ \rightarrow 0^+$ transition in $^{38}$Ar \cite{NNDC}. The fact that the 2174 keV line is different from the contaminant 2167 keV line can be seen in Fig. \ref{fig:2174}, via the difference in the location of their centroids from rather clean gates for both $^{38}$Ar and $^{41}$K.

As expected, this heavy-ion fusion-evaporation reaction strongly favored yrast and near-yrast states. Except for the 7841 keV state and the two states decaying into it, all the newly observed states decay only into the positive or negative parity sequence. For clarity of display, in Fig. \ref{fig:41K_lvl} the new states have been placed above the states they decay into. This should not be taken to imply any parity assignments to the newly observed states.

The high energy (and likely high-spin) newly observed states decay to the three previously known yrast levels of $13/2^+$, $15/2^+$, and $19/2^-$, confirming J$^{\pi}$ = $15/2^+$ for the 3899 keV level. Based on the presented facts, it appears that higher spin states can be observed only at the cost of high excitation energies and the newly observed states are likely to be higher than 13/2$^+$ and 19/2$^-$. Additionally, these observations are consistent with the Pauli exclusion principle. The two $f_{7/2}$ neutrons can couple to a maximum spin of 6 and if no neutron has been promoted to the $f_{7/2}$ orbital (a 0ph excitation), an additional spin of 3/2 (from the odd proton in the $d_{3/2}$ orbital) is added to give a maximum spin of $15/2^+$. If a valence proton is being promoted to $f_{7/2}$ (a 1ph excitation), then the maximum total spin for the three $f_{7/2}$ nucleons will be $19/2^-$. Additionally, promotion of a deeply buried $d_{5/2}$ proton to $d_{3/2}$ gives spin of $17/2^+$ for the 0ph configuration. The same procedure carried out for the 1ph configuration results in a maximum spin of $27/2^-$ (19/2 + 3/2 + 5/2). In conclusion, states involving the promotion of either one or two extra nucleons to the fp shell can be considered as candidates for the structure of the newly observed states.

The newly observed states do not exhibit long decay chains above them. Also, despite careful searches, none was observed to decay to states with spins below $13/2^+$ or $19/2^-$. It is unlikely that their spins are less than $13/2^+$ or $19/2^-$ or other decay paths would be available. In fact most spins are likely to be higher than $15/2^+$ and $19/2^-$. 

Now, we will see what can be learned experimentally about spins and parities from the present experiment. This is limited in different ways by the relative weakness of the new lines and the fact that polarizations and angular effects are relatively small. While the number of counts is not a limiting factor for many of these lines, smaller peak-to-background ratios mean that accuracy is limited by background determinations.

\begin{figure}[h!]
\includegraphics[width=\linewidth]{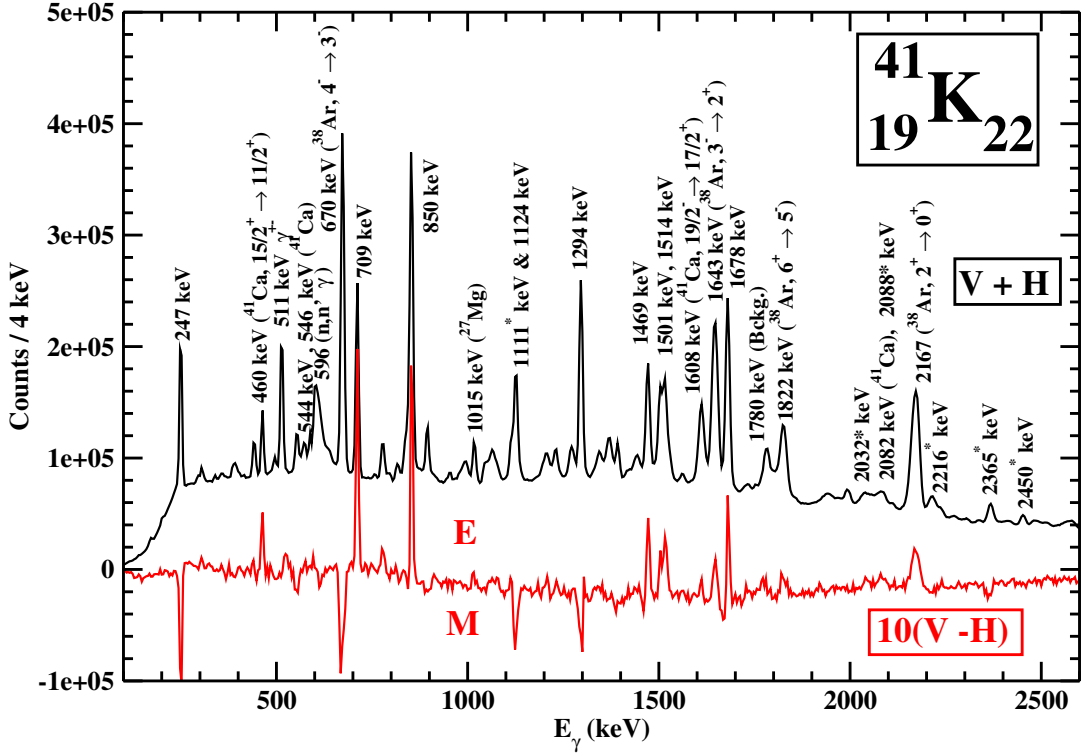}	
\caption{A comparison of the sum and difference of the gated spectra of Compton scattering events between vertical and horizontal pairs of HPGe crystals in each clover detector located near 90$^o$ relative to the beam direction. Scattering perpendicular to the beam is more probable for electric transitions (upward trending peaks) and parallel (downward trending peaks) for magnetic ones. All of the newly observed $^{41}$K transitions are marked with an asterisk (*).}
\label{fig:41K_pol}
\end{figure}

\begin{figure}[h!]
\includegraphics[width=\linewidth]{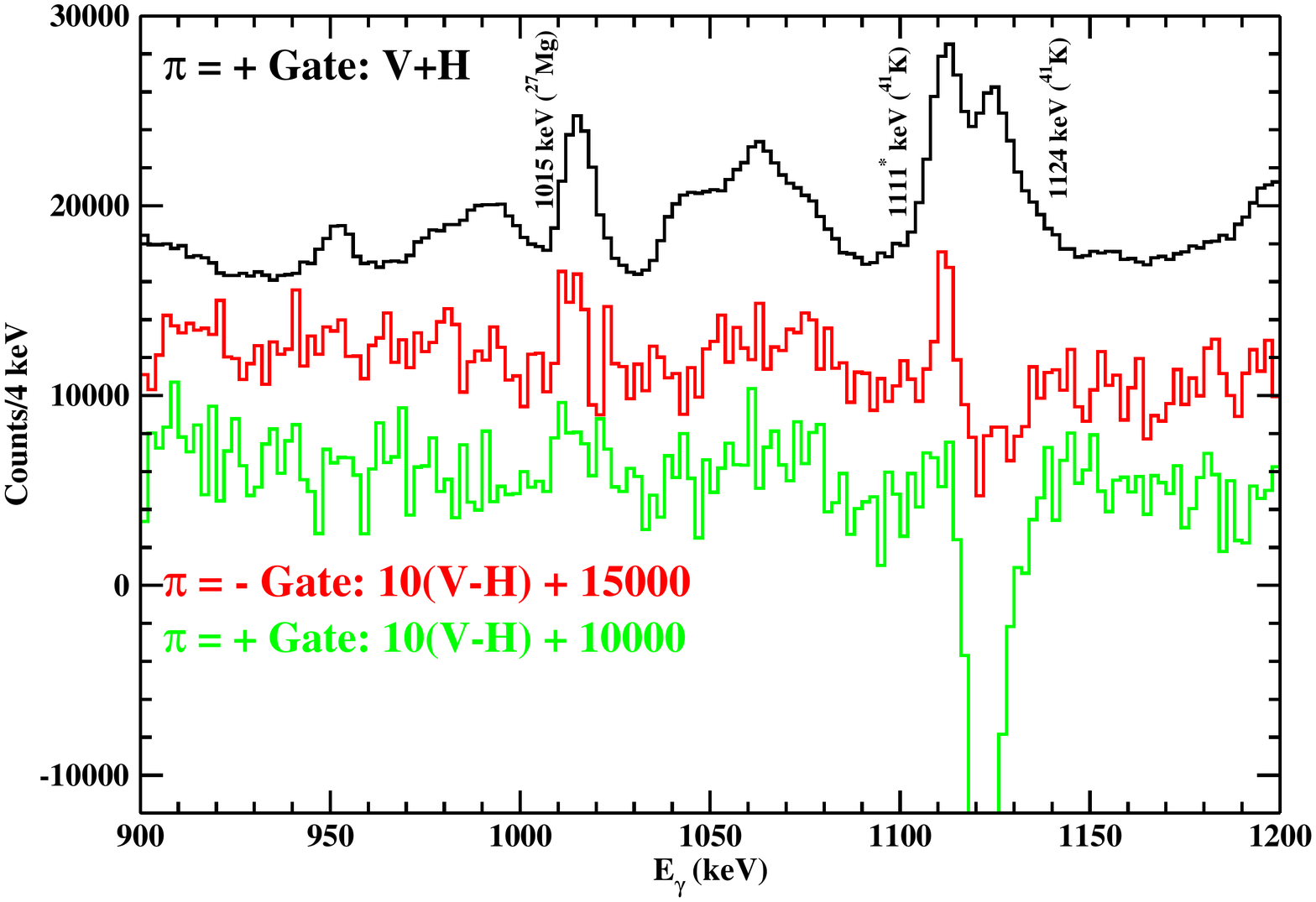}	
\caption{Spectra similar to that in Fig. \ref{fig:41K_pol}, but expanded to highlight the 1111 keV and 1124 keV $\gamma$ transitions in $^{41}$K. Separation of these close peaks in the difference spectra is enhanced by gates in the positive- and negative-parity decay chains. It should be noted that for the sum spectrum, only the positive parity gates were used to minimize the $^{38}$Ar background.}
\label{fig:41K_polB}
\end{figure}

Comparing the sum of Compton scattered perpendicular and parallel transitions to the beam direction with the difference between them provides the most reliable determination of the relative strength of electric to magnetic radiation. This comparison is shown in Figs. \ref{fig:41K_pol} and \ref{fig:41K_polB}. These spectra have been gated by most of the strong $^{41}$K lines. The polarization asymmetries agree well with the previous J$^\pi$ assignments for $^{41}$K, $^{41}$Ca, and $^{38}$Ar. Fig. \ref{fig:41K_polB} shows an expanded region around the 1111 keV-1124 keV doublet. As expected the 1124 keV line exhibits a magnetic behavior, while the newly observed 1111 keV line is dominated by electric emission. The fact that five newly observed higher-lying, higher-spin, states decay to the 3899 keV level strongly favors $15/2^+$ for its spin and effectively and rules out the other previously assigned possible spin of $11/2^+$. The $135^o$ + $45^o$ to $90^o$ R$_{\rm AD}$ values (See Table \ref{Tab:41K}) for the previously reported transitions are also consistent with their reported spin assignments. The remaining discussion will focus on what can be determined experimentally about spin and parity assignments to the newly observed, more weakly populated states.

The strongest new line is the 2365 keV decay to the 4984 keV, $19/2^-$, level. The polarization of this decay is clearly characteristic of magnetic dominance (See Fig. \ref{fig:41K_pol}). An M2 decay from a $23/2^+$ state is ruled out because it would imply a long lifetime leading to an absence of Doppler shifting like that of the 1294 keV line, while the 2365 keV one exhibits a Doppler shifting similar to most of the other lines. This leaves M1 decay, possibly mixed with E2. A $21/2^- \rightarrow 19/2^-$ decay gives an excellent fit to the angular distribution with a mixing ratio $\delta$ of about -0.8. Other possibilities of $17/2^-$ and $19/2^-$ are much less consistent with the R$_{\rm AD}$ value. So we assign $21/2^-$ for the 7348 keV state. 

The 1111 keV transition decays to the 7348 keV, parent state of the 2365 keV transition, and its polarization is clearly of electric type. This gives three possibilities for its parent state, $19/2^+$, $23/2^+$, or $25/2^-$. The angular distribution ratio agrees well with a pure E1 decay but not for a J to J one, nor an E2 decay. The lower spin of $19/2^+$ is much less likely since it could decay to several lower energy states, but does not. Therefore we assign $23/2^+$ to the 8460 keV state.

Another relatively strong line among the newly observed ones is the 2174 keV transition. We have already commented on care taken with this line so close to the strong $^{38}$Ar 2167 keV one which contaminates some of the $^{41}$K gates. Unfortunately the strong electric type polarization of the 2167 keV line still dominates and we cannot conclude anything about the parity of the 6073 keV state. The R$_{\rm AD}$ value implies a nearly pure dipole transition and $13/2$ or $17/2$ for the 6073 keV parent state. Given that a $13/2$ state of either parity could decay to many different states below it and none of whose decays have been observed, $17/2$ is by far the most likely possibility for the 6073 keV level.

There is no reliable indication of a polarization asymmetry of the 949 keV transition, but its angular distribution asymmetry, R$_{\rm AD}$, is consistent with a $17/2 \rightarrow 19/2^-$ or ($21/2 \rightarrow 19/2^-$) decay with a mixing ratio $\delta$ of about +0.21 (-0.27). These mixing ratios are large enough to strongly favor an M1/E2 decay over an E1/M2 one. This implies a spin-parity of $17/2^-$ or $21/2^-$ for the newly observed 5934 keV level.

The R$_{\rm AD}$ of the 2216 keV line is consistent with a $17/2 \rightarrow 19/2^-$ ($21/2 \rightarrow 19/2^-$) decay with a mixing ratio $\delta$ of about -0.34 (+0.25). As mentioned above, the nonzero mixing ratios imply an M1/E2 transition which is consistent with its M type polarization asymmetry, implying a J$^\pi$ assignment of $17/2^-$ or $21/2^-$ for the newly observed 7200 keV level.

The R$_{\rm AD}$ of the 2580 keV line is almost identical with that of the 2216 keV transition, implying a $\Delta J = \pm 1$ decay with a significant mixing ratio. There is no discernible polarization asymmetry but a mixed M1/E2 transition is much more likely than an E1/M2 one, strongly suggesting $17/2^+$ or $13/2^+$ for the parent 6478 keV state, with the higher spin strongly favored from yrast considerations. 

The 2450 keV line also shows no clear indication of a polarization asymmetry even though it is relatively strong among the newly observed decays. Its R$_{\rm AD}$ is only consistent with a $15/2 \rightarrow 13/2^+$ decay with a relatively small mixing ratio of $\delta \approx -0.11$. This implies $15/2$ for the 5224 keV state without a parity assignment.

The R$_{\rm AD}$ of the 2683 keV line is almost identical to that of the 949 keV transition implying a $\Delta J = \pm 1$ decay with a significant mixing ratio. Both the large mixing ratio and a likely magnetic decay polarization asymmetry imply an M1/E2 decay and $11/2^+$ or $15/2^+$ for the 5459 keV state, with the lower spin essentially ruled out from yrast considerations.

The polarization asymmetry for the 2774 keV line suggests a predominantly electric character. The R$_{\rm AD}$ value of the 2774 keV line gives the best fit for a nearly pure E2 transition with $\delta \approx -0.03$. This coupled with the electric tendency would imply $23/2^-$ for the 7758 keV state. The R$_{\rm AD}$ is also consistent with $19/2^+$, but the observation of the 4597 keV line feeding the 7758 keV state from a level 2 MeV above the neutron separation energy effectively rules it out. Together these observations lead to a $J^\pi$ assignment of $23/2^-$ for the 7758 keV level.

The 7841 keV level is particularly interesting because it decays to both the positive and negative decay sequences. In turn, two of the highest energy states observed decay to it and to no state below it. These properties essentially limit the spin of the 7841 keV state to as high a spin as possible. On the other hand, its decay branches to states of spin $J$ = $17/2$ and $J^\pi$ = $19/2^-$ constrain its spin to $\leq$ 21/2. Although its 1770 keV decay branch is only partially resolved from a 1780 keV contaminant line of comparable intensity, its polarization clearly implies electric type. The polarization asymmetry of its 2855 keV decay to the known $19/2^-$ suggest magnetic type, but is not conclusive. Taken together, the evidence is very suggestive that J$^\pi$ = (21/2$^+$) for the 7841 keV state and (17/2$^+$) for the 6073 keV state.

\subsection{$^{41}$Ca}
\begin{figure}[h!]
\includegraphics[width=\linewidth]{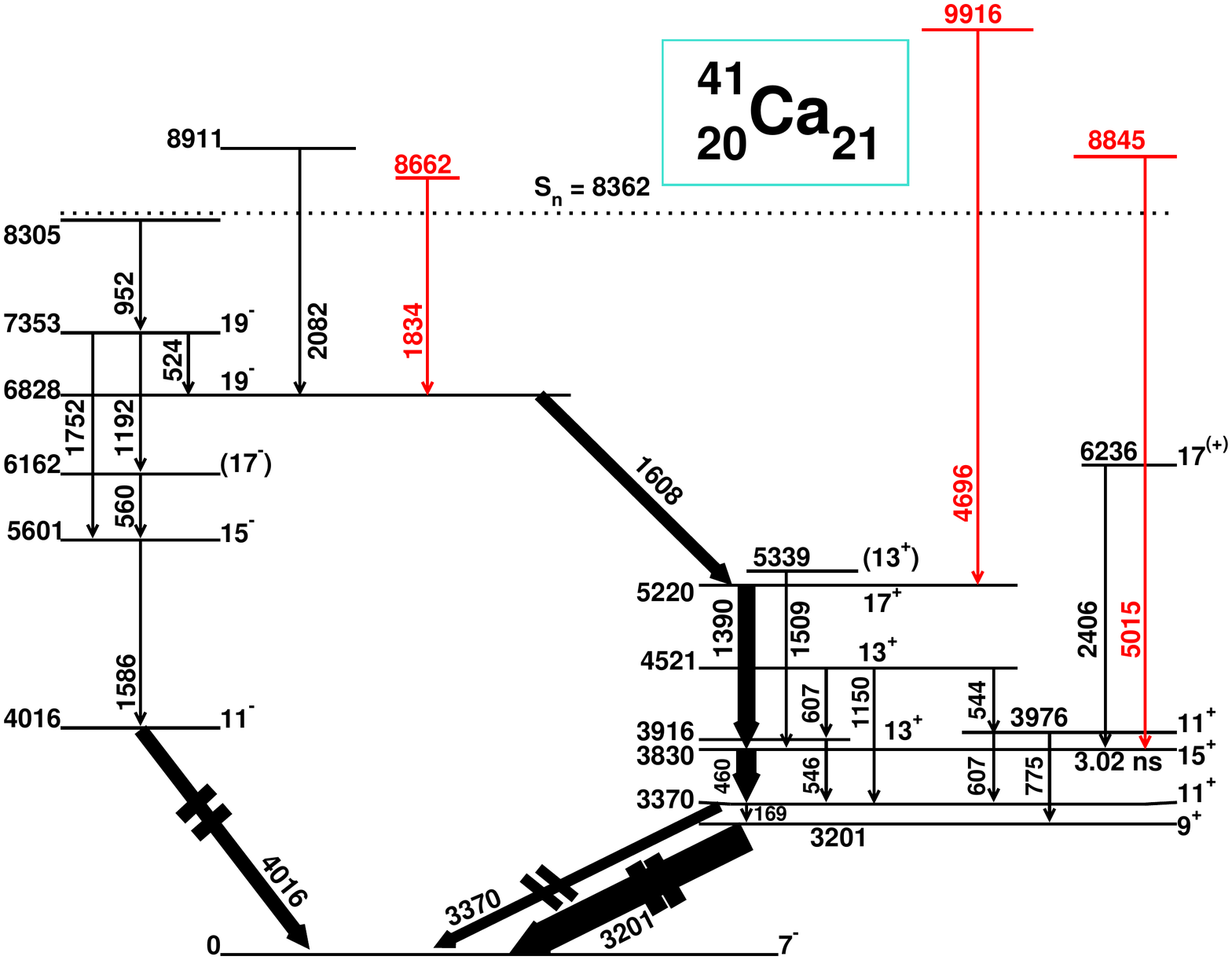}
\caption{States (with spins given as 2J) and $\gamma$ decays observed in the present experiment in $^{41}$Ca (with newly observed ones in red). The dashes through the three lowest transitions in the decay scheme indicate that the length of the arrows is not proportional to the energy of those transitions. The neutron separation energy, S$_n$, is represented by a horizontal dotted line. The newly observed states are placed above the previously known decay sequences to which they decay (this does not imply that they have the same parities as them). The widths of the decay arrows give an indication of their intensities. Note that the lifetime of the 3830 keV state comes from Ref. \cite{NNDC}.}	
\label{fig:41Ca_lvl}
\end{figure}

\begin{figure}[h!]
 \centering
 \includegraphics[width=\linewidth]{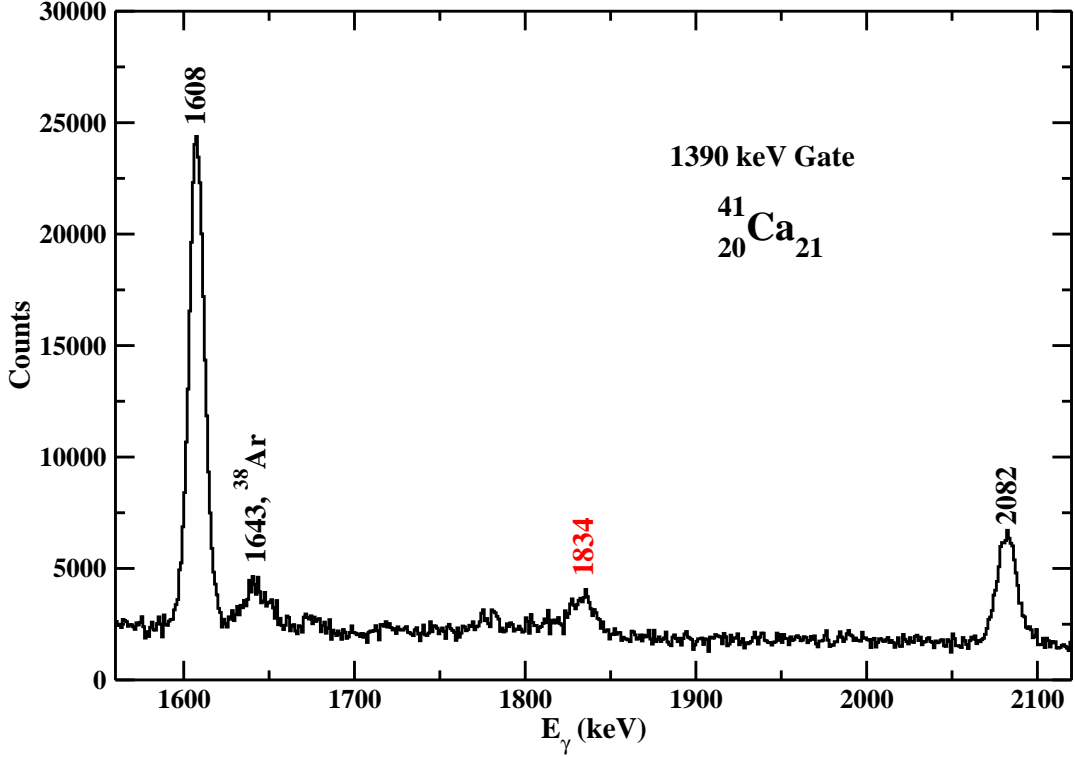}
\caption{A portion of the $\gamma$ spectrum in coincidence with the 1390 keV previously reported transition showing the newly observed 1834 keV transition in $^{41}$Ca.}
\label{1389_4014_41Ca}
\end{figure}

\begin{figure}[h!]
\centering
\includegraphics[width=\linewidth]{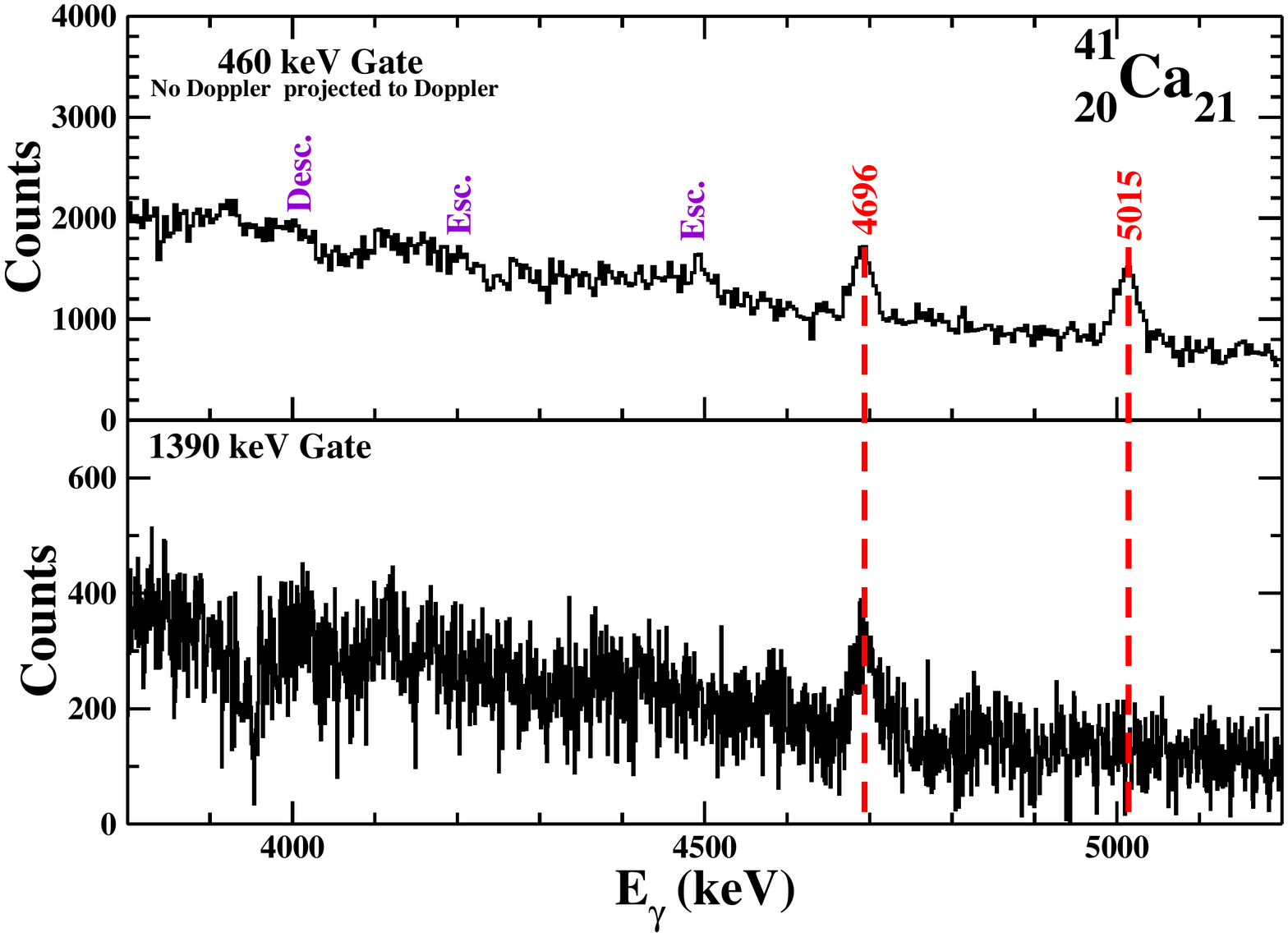}	
\caption{A portion of the Doppler-corrected $\gamma$ spectrum in coincidence with the unDoppler-corrected 460 keV previously known transition (top) and the Doppler-corrected 1390 keV one (bottom) in $^{41}$Ca showing the newly observed rather high energy decay lines (in red). The single and double lepton escape peaks are also labeled.}
\label{fig:41Ca_460gate}
\end{figure}

The reaction also produced a good amount of data for $^{41}$Ca, always of interest so close to the shell boundary, so we have explored it for new transitions and states for completeness. Experimentally this is more challenging because the lowest transitions have rather high energies and, thus, low efficiencies for $\gamma$ detection. The decay scheme is shown in Fig. \ref{fig:41Ca_lvl}. In spite of the many previous studies of $^{41}$Ca, three newly observed states and their decays were identified. All three states lie above the neutron decay threshold. Several key coincidence spectra, with the 460 keV and 1390 keV $\gamma$ rays, illustrating the newly observed transitions are shown in Figs. \ref{1389_4014_41Ca} and \ref{fig:41Ca_460gate}. Since no other decays of these three states were observed, their spins are likely to be 1 or 2 $\hbar$ above those of the states they decay to.

\section{Discussion}

Possible spins in $^{41}$K (with one proton hole and two additional neutrons relative to the doubly magic $^{40}$Ca nucleus) are rather limited without promoting additional nucleons to the $fp$ shell. Recoupling of the the $\nu f_{7/2}$ pair from $0 $ (the predominant g.s. configuration) to the maximum consistent with the Pauli principle of $6 $ gives the lower $\pi = +$ states which form the g.s. decay sequence up to $15/2^+$. After the g.s., the yrast sequence switches to the $J^{\pi}$ = 7/2$^-$ (1ph) state because it appears that promoting the odd $d_{3/2}$ proton to $f_{7/2}$ is somewhat less energetically expensive than just recoupling $\nu f_{7/2}$ pair to $2 $ because of the pairing gap. The lower spin negative-parity states could be formed by either promotion of a proton or another neutron to the $f_{7/2}$ orbital or by a mix of both. Promoting a proton to the $f_{7/2}$ shell (1ph) and recoupling the remaining $\pi d_{3/2}$ pair to $2 $ would give J$^{\pi} = 23/2^-$ and even slightly more spin is available from promotion of a deeply buried $d_{5/2}$ proton. These higher spin 0ph and 1ph states are likely to be in the same energy range as 2ph and 3ph configurations. Together, all of these types of higher spin configurations are likely to account for the wealth of newly observed states.

\begin{figure}[htbp]
\centering
\subfloat{\includegraphics[width=0.45\textwidth]{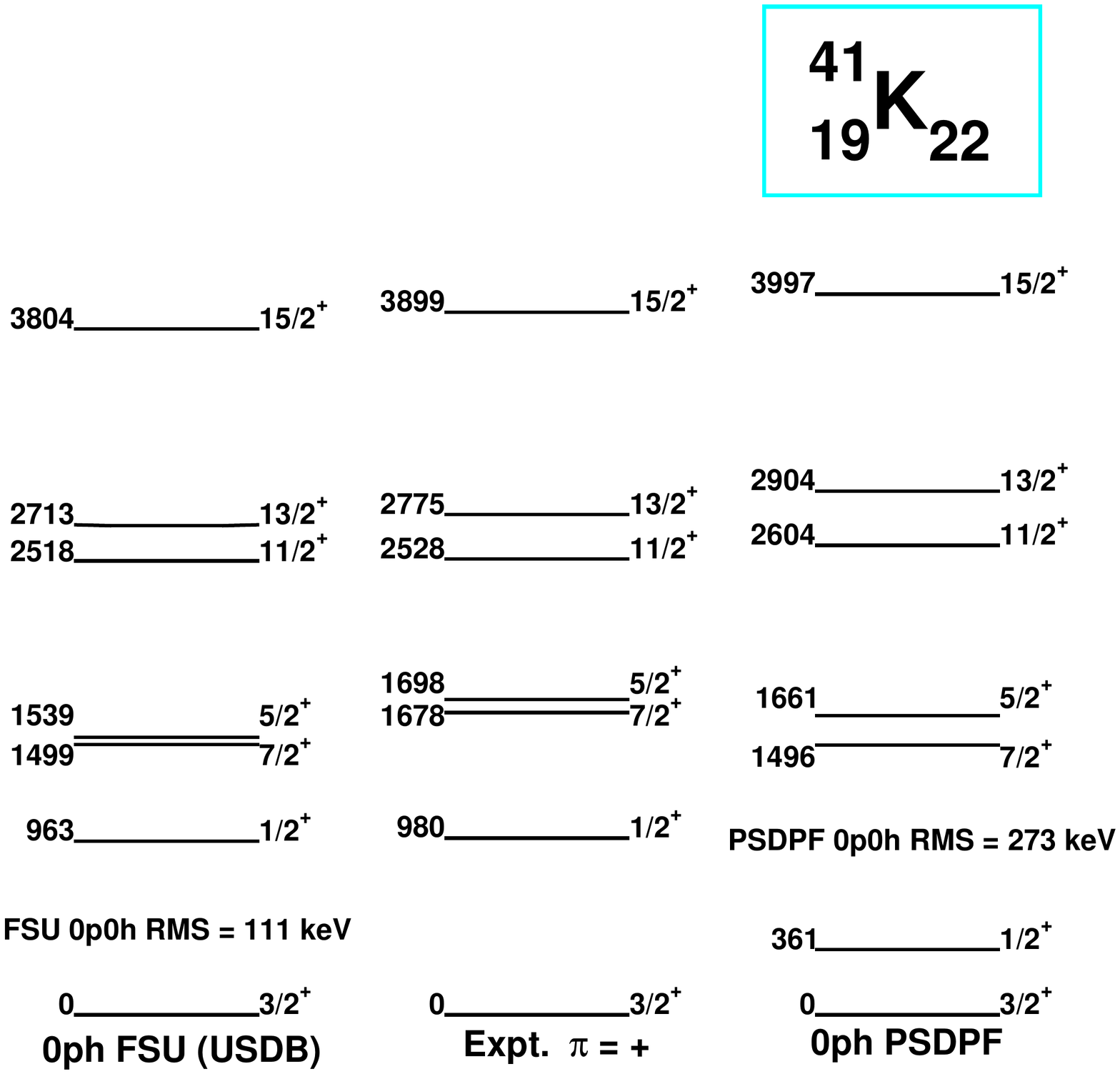}}
\subfloat {\includegraphics[width=0.45\textwidth]{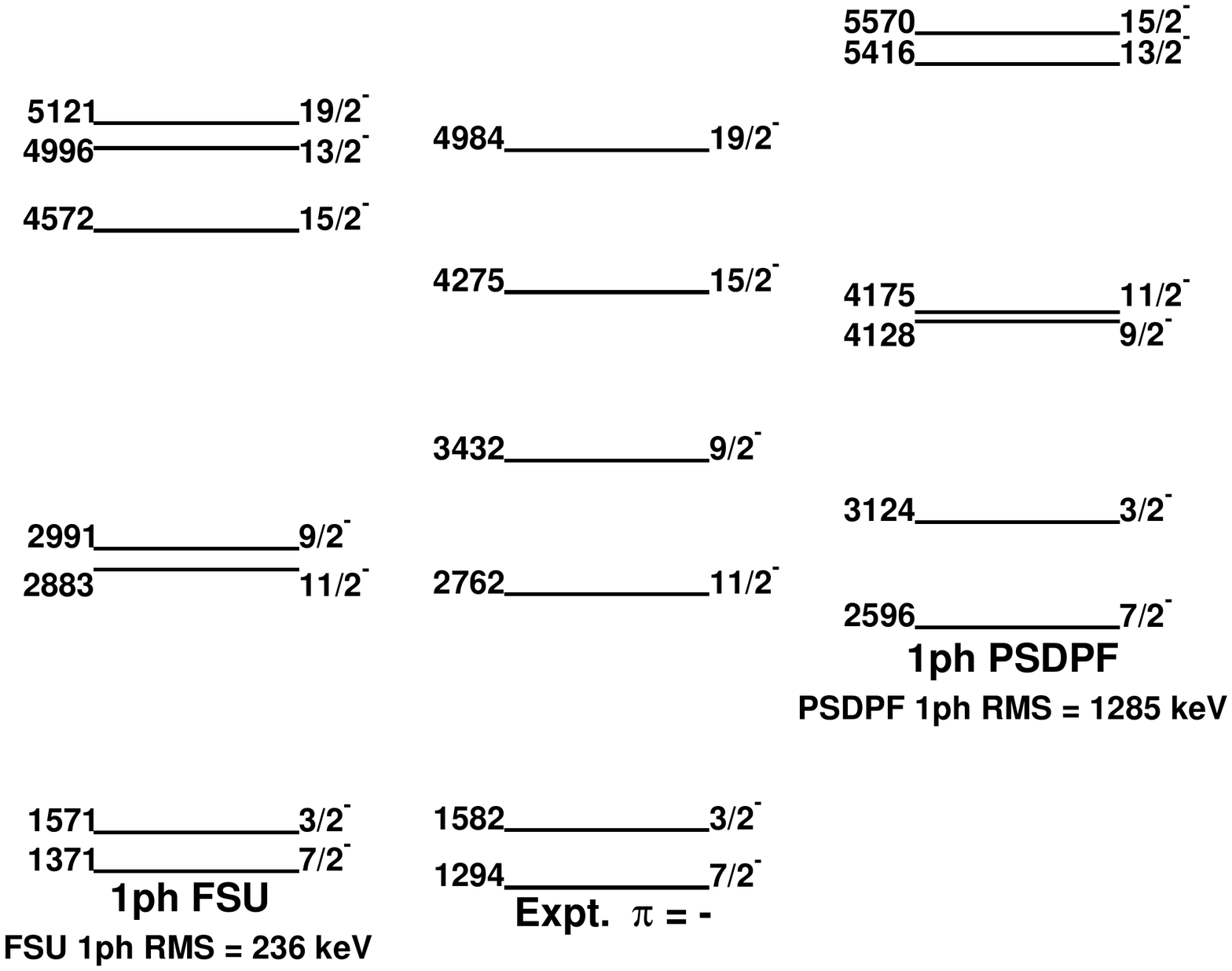}}
\caption{A comparison of the low lying experimental states in $^{41}$K with the FSU \cite{lubna_evolution_2020,lubna_structure_2019} and PSDPF shell model interaction calculations illustrating the improved agreement with the FSU interaction. The 0ph FSU interaction calculations are from the USDB \cite{brown_2006} interaction.}
\label{fig:41K_PSDPF}
\end{figure}

A comparison between the 0ph and 1ph experimental, FSU shell model interaction \cite{lubna_structure_2019,lubna_evolution_2020}, and PSDPF \cite{psdpf} shell model interaction levels is shown in Fig. \ref{fig:41K_PSDPF}. Note, the FSU interaction 0ph calculations are the same as those from the USDB \cite{brown_2006} interaction. The FSU (USDB) interaction 0ph energies have slightly better agreement with the experimental ones than the PSDPF interaction ones. For the 1ph excitations, the FSU interaction has quite a bit better agreement with the experimental energies compared to the PSDPF interaction results. 

\begin{figure}[h!]
\includegraphics[width=\linewidth]{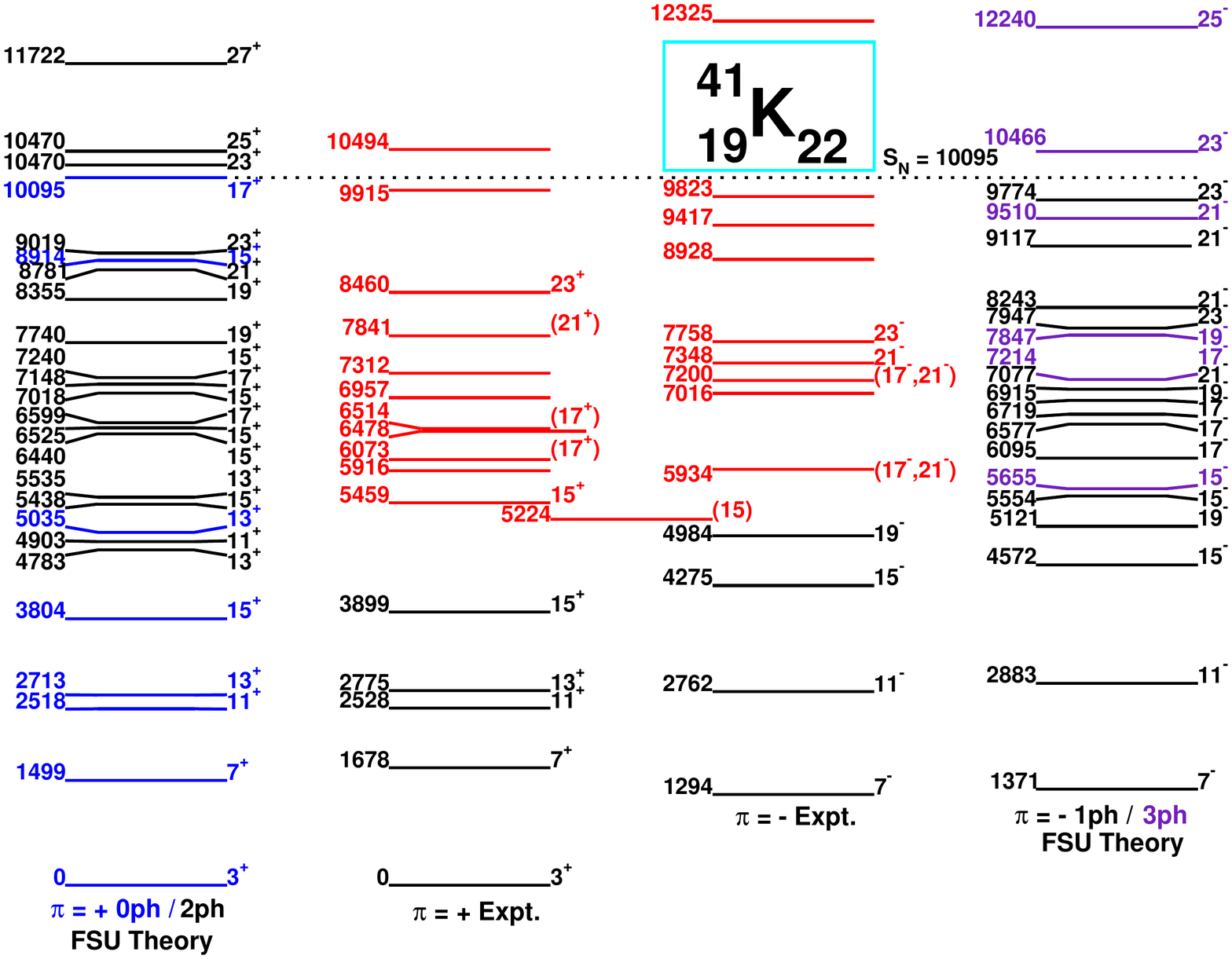}	
\caption{A comparison of the higher spin experimentally observed states in $^{41}$K with 0ph (in blue), 1ph (in black), 2ph (in black), and 3ph (in purple) predictions of the FSU cross-shell interaction for the indicated excitations. Spins are given as 2J and the newly observed levels, in red. Comparisons are discussed in the text.}
\label{fig:41K_FSUthry}
\end{figure}

The yrast and near-yrast states predicted by the FSU interaction for 0ph, 1ph, 2ph, and 3ph configurations in $^{41}$K are shown in Fig. \ref{fig:41K_FSUthry}. The FSU shell model calculations for 0ph and 1ph reproduce the excitation energies of the previously assigned yrast and near-yrast states well in both the positive- and negative-parity decay sequences with root-mean-square (RMS) differences of about 100 and 250 keV, respectively. The full $fp$ shell is open for these calculations but its occupancy is limited to 2 neutrons for the 0ph case. For the $3/2^+$ g.s. these 2 neutrons lie predominantly (80\%) in the $f_{7/2}$ orbital, an occupancy which increases to 95\% for the $7/2^+$ state and is almost zero for the higher spins. The highest spin possible without allowing any more nucleons above the $sd$ shell is $17/2^+$, formed by promotion of a deeply buried $\pi d_{5/2}$ to $\pi d_{3/2}$ at a total energy ``cost" of 10.095 MeV. Such a state is unlikely to be observed in this reaction since higher spin 1ph and 2ph configurations will be less energetically expensive.



Clearly higher spin positive-parity states within the range of this experiment will involve 2ph excitations, note that the FSU interaction was not directly fitted to any 2ph states although some reasonable agreement has been reported in Refs. \cite{PhysRevC.100.014310} and \cite{lubna_structure_2019}. The lowest calculated 2ph state lies at 2554 keV, about twice that of the lowest 1ph one. However, it is a low spin $3/2^+$ level. The lowest 2ph $15/2^+$ state at 5438 keV lies about 1600 keV above the lowest 0ph state. The newly observed 5459 keV level provides an excellent energy match for this, although the 5224 keV state cannot be ruled out and there are indications that the 2ph energies may be over predicted. Two more $15/2^+$ levels are predicted at 6440 and 6525 keV which could match other new states if these energies are predicted a few hundred keV too high. The lowest predicted 2ph $17/2^+$ state lies at 6599 keV compared to the only $17/2^+$ one for 0ph at 10095 keV, so the positive-parity yrast line switches from 0ph to 2ph between 15/2$^+$ and 17/2$^+$. The newly observed level at 6478 keV provides the best energy match, but the 6073 keV ($17/2^+$) candidate cannot be ruled out. In fact when looking for further correspondences with experiment, it appears likely that the higher spin 2ph states may be predicted too high by good fractions of an MeV. In this interpretation the 6073 - 6599 keV experiment - theory identification could be followed by 6478 - 7148 keV for another $17/2^+$ state and 8781 - 7841 keV for $21/2^+$, and 8460 - 9019 keV for $23/2^+$. These uncertainties point to more work needed to improve the reliability of theory.

The lowest negative-parity states must involve 1ph configurations. As a reminder, for these calculations the full $fp$ shell was kept open. The lowest 1ph state is the $7/2^-$ one predicted at 1371 keV and observed nearby at 1294 keV. Comparing proton to neutron excitation, the calculated occupancies are close to two neutrons and one proton in the $fp$ shell for all the 1ph yrast states. That is, they are almost pure proton excitations. Some of the other lower spin states range up to roughly half-and-half proton and neutron excitations. Presumably proton excitation is favored because it does not require breaking a nucleon pair. The highest previously assigned spin is $19/2^-$ at 4984 keV, but no $17/2^-$ state has been previously reported. The FSU 1ph calculation predicts the lowest $19/2^-$ level in good agreement at 5121 keV. It also accounts for the ``missing" $17/2^-$ level because the lowest one is predicted an MeV higher than the $19/2^-$ state at 6095 keV. The newly observed 5934 keV state is a good candidate for that. The lowest predicted 1ph $21/2^-$ level is predicted at 7077 keV and we have observed several states which are good candidates. The lowest predicted $23/2^-$ state at 7947 keV provides a good candidate for the newly observed 7758 keV level. The lowest predicted $25/2^-$ 1ph state lies at 13778 keV, about 1.5 MeV above the highest energy state observed in this experiment. It appears that 1ph configurations cannot account for all the states observed decaying into known negative-parity states. This leads to the question of what about 3ph structures.

Given the considerations above for the calculated energies for 2ph states, the 3ph excitations should be cautiously considered. The lowest predicted one is $7/2^-$ at 3513 keV. However, the 3ph energies only drop below the 1ph ones for $25/2^-$ at 12240 keV, close to the highest-lying state observed in the present experiment. It is likely that the energies predicted for 3ph configurations are overpredicted, but this uncertainty is becoming too large to make useful comparisons with experiment. 

We can also take a quick look at where the newly observed $^{41}$Ca states would fit. The lowest $\pi = +$ states should be 1ph excitations. The maximum spin of the configuration $(\nu d_{3/2})^{-1} \otimes (\nu f_{7/2})^2$ (relative to the $^{41}$Ca (g.s.)) is $15/2^+$ which is one of the most strongly populated states and, therefore, yrast. Due to the Pauli principle, promotion of a proton instead gives the higher spin of $17/2^+$ with $\pi f_{7/2} \otimes (\pi d_3/2)^{-1} \otimes \nu f_{7/2}$ which must be the dominant configuration of the next yrast state. In fact this 5220 keV state was the most strongly populated one in an earlier $^{39}$K($\alpha$,d) reaction \cite {nann_1975} and interpreted as this configuration coupled to the maximum spin.

Above spin 17/2 the yrast line passes to 2ph excitations. Configurations such as $ (\pi f_{7/2})^2_6 \otimes (\pi d_{3/2})^{-2}_0 \otimes \nu f_{7/2}$; $(\nu f_{7/2})^3_{15/2} \otimes (\nu d_{3/2})^2_2$; and a mixed proton-neutron excitation can give states of $19/2^-$. Spins up to 2 units more can be formed by coupling the $d_{3/2}$ holes to spins 1 or 2. These latter configurations could account for th 8911 keV and newly observed 8662 keV states. Because of their high energies and relatively strong populations, the newly observed 8845 keV and 9916 keV states likely have J$^\pi$ values of $19/2^+$ and $21/2^+$ and give us the first sightings of 3ph excitations in $^{41}$Ca.

\section{Summary}
Gamma decay correlations following the fusion of $^{18}$O with $^{26}$Mg at 50 MeV have provided interesting new information on the structure of $^{41}$K and $^{41}$Ca near the neutron and proton $sd$-$fp$ shell gap. This work extended the decay schemes of $^{41}$K and $^{41}$Ca up to 12325 keV and 9916 keV, respectively, both above their neutron emission thresholds. The newly observed high-spin states in $^{41}$K decayed only into the highest spin previously observed states. The reaction strength was surprisingly spread out over a number of states with no clear continuation of the yrast sequences. Additionally, three previously unreported $\gamma$ transitions were identified in 41Ca, two of which decayed into the positive-parity yrast levels. Analyses of angular distributions and polarization spectra were used to tentatively assign spins and parities for a number of $^{41}$K states, limited by the weaker population of the newly observed states. 

Higher-spin structures in $^{41}$K were calculated using the recently determined FSU cross-shell interaction. The agreement between calculated 0ph and 1ph energies with those of the previously assigned yrast states is very good, and good candidates for several of the newly observed states are proposed. However, most of the newly observed states likely have higher spins and/or lower energies than can be explained by 0ph and 1ph excitations. Reasonable matches for the new levels decaying into the known yrast positive-parity states can be found among the 2ph calculations if those energies are overpredicted by a few hundred keV to an MeV. 

Overall, this study has revealed many more relatively high-spin excitations around the $sd$-$fp$ shell boundary and the value of precision $\gamma$ spectroscopy even well above the neutron decay thresholds. It has given a general picture of how multiparticle cross-shell excitations can build up spin at the lowest energy cost. This study and other such ones can paint the way to build an even more comprehensive multishell microscopic theory.

%
%

\begin{acknowledgements}
This material is based upon work supported by the U.S. National Science Foundation under Grant No. PHY-1401574 (FSU), PHY-1712953 (FSU), and PHY-2012522 (FSU), the Stewardship Science Academic Alliance through the Centaur Center of Excellence under Grant No. NA0003841 (FSU), the U.S. Department of Energy, office of Science, under Awards No. DE-SC-0009883 (FSU) and DE-AC05-00OR22725 (ORNL). 
\end{acknowledgements}

\end{document}